\documentclass[%
 reprint,
 amsmath,amssymb,
 aps,
]{revtex4-2}

\usepackage{graphicx}% Include figure files
\usepackage{dcolumn}% Align table columns on decimal point
\usepackage{bm}% bold math
\newcommand{\gdot}{\dot{\gamma}}

\begin{document}

\preprint{APS/123-QED}

\title{Reduced stress propagation leads to increased mechanical failure resistance in auxetic materials}% Force line breaks with \\
%\thanks{A footnote to the article title}%

\author{Suzanne M. Fielding}%
\affiliation{Department of Physics, Durham University, Science Laboratories,
  South Road, Durham DH1 3LE, UK 
}%

\date{\today}% It is always \today, today,
             %  but any date may be explicitly specified

\begin{abstract}
Materials with negative Poisson ratio have the counter-intuitive property of expanding laterally when they are stretched longitudinally. They are accordingly termed auxetic, from the Greek auxesis meaning to increase. Experimental studies have demonstrated auxetic materials to have superior material properties, compared with conventional ones. These include synclastic curvature, increased acoustic absorption, increased resilience to material fatigue, and increased resistance to mechanical failure. Until now, the latter observations have remained poorly understood theoretically. With this motivation, the contributions of this work are twofold. First, we elucidate analytically the way in which stress propagates spatially across a material following a localised plastic failure event, finding a significantly reduced stress propagation in auxetic materials compared with conventional ones. In this way, a plastic failure event occurring in one part of a material has a reduced tendency to trigger knock-on plastic events in neighbouring regions. Second, via the numerical simulation of a lattice elastoplastic model, we demonstrate  a key consequence of this reduced stress propagation to be an increased resistance to mechanical failure. This is seen not only via an increase in the externally measured yield strain, but also via a decreased tendency for plastic damage to percolate internally across a sample in catastrophic system-spanning clusters.
\end{abstract}

%\keywords{Suggested keywords}%Use showkeys class option if keyword
                              %display desired
\maketitle

When a material is stretched longitudinally, its ratio of lateral contraction to longitudinal stretch defines its Poisson ratio $\nu$. For an isotropic material in $d$ spatial dimensions, $\nu$ varies between $+1/(d-1)$ in an incompressible material, for which the ratio $G/K=0$, and $-1$ in the limit of perfect compressibility, $G/K\to \infty$. Here $G$ and $K$ are respectively the material's  shear and  bulk modulus. Materials with negative Poisson ratio  have the counter-intuitive property of expanding laterally when stretched, and are termed auxetic~\cite{lakes1987foam,evans1991molecular,evans1991auxetic} from the Greek auxesis meaning increase. Important categories of auxetics include solid cellular  foams~\cite{lakes1987foam} or honeycombs~\cite{gaspar2005novel,gibson1982mechanics}, microporous polymers~\cite{caddock1989microporous,evans1989microporous}, composites~\cite{clarke1994negative,alderson2005make}, some biomaterials such as skin~\cite{veronda1970mechanical} and bone~\cite{williams1982properties}, and some crystalline materials~\cite{gunton1972young,baughman1998negative}. They have also been created from random fibre networks~\cite{mirzaali2019auxeticity}, often by bond pruning~\cite{bonneau2021geometric,reid2018auxetic}, and via elastic buckling instabilities~\cite{bertoldi2010negative,babaee20133d,domaschke2019random,shim2013harnessing} and origami~\cite{lv2014origami}. At the level of internal micro- or meso-scopic structuring, auxetics are often based on chiral sub-units~\cite{alderson2010elastic}, rotating rigid sub-units~\cite{grima2000auxetic,grima2005auxetic,gatt2015hierarchical}, or re-entrant  sub-units with negative internal angles~\cite{evans2000auxetic}.   For reviews, see ~\cite{yang2004review,prawoto2012seeing,lakes2017negative,lakes1993advances,alderson2007auxetic,evans2000auxetic,kolken2017auxetic,ren2018auxetic, saxena2016three}.

Potential applications of auxetics have been proposed in multiple arenas: in biomedical devices such as stents~\cite{ali2014auxetic}, skin grafts~\cite{gatt2015hierarchical} and artificial blood vessels~\cite{evans2000auxetic}; in protective sports pads, helmets, mats and shoes~\cite{sanami2014auxetic}; in molecular sieves~\cite{alderson2005make} and defouling applications~\cite{alderson2000membrane}; in  textiles~\cite{sloan2011helical}; and as smart sensors~\cite{grima2013smart}. Such widespread potential uses stem from the superior mechanical properties of auxetic materials compared with conventional ones. These include  synclastic curvature~\cite{lakes1987foam}, increased acoustic absorption~\cite{chen1996micromechanical,scarpa2006dynamic}, increased indentation resistance~\cite{lakes1993indentability,chan1998indentation},  increased resistance to material fatigue~\cite{bezazi2009tensile} and -- of primary concern to this work -- increased resistance to mechanical failure~\cite{hall2008sign,yang2017fracture,lakes1987foam,donoghue2009fracture,evans1990tailoring,choi1996fracture,bezazi2009tensile}.

Experimentally, the increased  resistance of auxetics to mechanical failure has been demonstrated in several studies. In carbon nanosheets, density-normalized sheet toughness, strength and modulus all increased at negative Poisson ratio~\cite{hall2008sign}. Auxetic Kevlar composites showed an improved fracture toughness  and a  reduction in damage area during impact testing, compared with conventional ones~\cite{yang2017fracture}.  An auxetic laminate in tension required more energy to propagate a crack than its conventional counterpart, with less notch sensitivity~\cite{donoghue2009fracture}.  Several studies have demonstrated an enhanced failure resistance of auxetic solid foams~\cite{choi1996fracture,bezazi2009tensile,lakes1987foam}, with stress-strain curves showing auxetic foams to survive to higher strains before collapse compared with conventional ones.
 In an indentation test~\cite{lakes1993indentability}, an auxetic foam yielded at higher stress, and with a reduced area of damage. 

Despite these observations, the increased mechanical failure resistance of auxetic materials  remains poorly understood theoretically. With this motivation, the key contributions of this work are twofold. First, we elucidate theoretically the way in which stress propagates through a material, whether conventional~\cite{picard2004elastic} or auxetic,  following a localised plastic event. We find a significantly reduced stress propagation in the auxetic case, such that a local plastic failure in one part of a material has a reduced tendency to trigger knock-on failure events in neighbouring regions. Second, we demonstrate an important consequence of this reduced stress propagation to be an increased resistance to mechanical failure.

\section*{Theoretical model}
\label{sec:model}

For definiteness, we cast much of the discussion that follows in the language of $d=2$ dimensional random fibre networks. However, we anticipate the basic model ingredients that we shall incorporate to serve as a minimal model for the mechanical failure of auxetics more generally. Indeed, we adopt a mesoscopic elastoplastic modelling approach related to that used to study yielding and flow in  a broad array of amorphous solids and complex fluids such as dense foams, emulsions, colloids, granular materials and metallic glasses~\cite{nicolas2018deformation}. In the spirit of these existing elastoplastic models, we coarse-grain a solid auxetic material at the mesoscopic level of a few bonds in a random fibre network, or a few cells in a solid foam or honeycomb, etc. Each such mesoscopic collection of bonds, cells, etc. is assumed large enough to allow the definition of local strain and stress variables, and is represented as a single elastoplastic element.  Local plastic failure events at the scale of these individual elements are then spatially coupled by elastic stress propagation at the global continuum level.

Compared with existing elastoplastic models, ours has two additional features.  First, when any elastoplastic element yields, it does not then reform (to model a T1 rearrangement of droplets in a flowing emulsion, say), but permanently fails, to model a localised `breakage' event in (e.g.) a solid foam or network. Second, whereas existing models typically consider incompressible materials, ours is generalised to model the full range of material types from incompressible to maximally auxetic. Indeed (in the auxetic case), we assume auxetic behaviour both on the mesoscopic lengthscale of the individual elastoplastic elements, as well as on the global continuum lengthscale over which stress propagation is computed.

On a square lattice of $N\times N$ sites, we consider a single elastoplastic element on each site. Site $(m,n)$ has position vector $r_i=(m\hat{\bf x}, n\hat{\bf y})$, with the lattice spacing defining our length unit.  As a function of space $r_{i}$ and time $t$, we define a strain relative to a state of undeformed equilibrium 
\begin{equation}
\label{eqn:strain}
    \epsilon_{ij}(r_i,t)=\tfrac{1}{2}\left(\partial_iu_j+\partial_ju_i\right),
\end{equation}
given a displacement field $u_i(r_i,t)$.
Associated with this strain is an elastoplastic stress $\sigma_{ij}$. At the level of linear elasticity, we write
\begin{equation}
\label{eqn:stress1}
\sigma_{ij}(r_i,t)=2\mu\epsilon_{ij}+\lambda\epsilon_{ll}\delta_{ij},
\end{equation}
with Lam\'e coefficients $\mu$ and $\lambda$. To relate these cofficients to the shear and bulk moduli, $G$ and $K$, we write the stress as a sum of contributions from  pure shear and isotropic compression:
%
%\begin{equation}
%\epsilon_{ij}=\left(\epsilon_{ij}-\frac{1}%{d}\epsilon_{ll}\delta_{ij}\right) + \frac{1}{d}\epsilon_{ll}\delta_{ij},
%\end{equation}
%
%and the stress correspondingly as
%
\begin{equation}
\label{eqn:stress2}
\sigma_{ij}=2G\left(\epsilon_{ij}-\frac{1}{d}\epsilon_{ll}\delta_{ij}\right) + K\epsilon_{ll}\delta_{ij}.
\end{equation}
Comparing Eqns.~\ref{eqn:stress1} and~\ref{eqn:stress2}, we then identify 
\begin{equation}
\label{eqn:LameElastic}
    \mu=G\;\;{\rm and}\;\; \lambda=-\frac{2G}{d}+K.
\end{equation}
%
%and
%\begin{equation}
%\label{eqn:LameElastic2}
%\end{equation}
%
The Poisson ratio
\begin{equation}
\label{eqn:Poisson}
\nu=\frac{dK-2G}{d(d-1)K+2G}.
\end{equation}
Of the five material constants $\mu,\lambda,G,K$ and $\nu$, therefore, we need specify only two, because the other three are determined via~\ref{eqn:LameElastic} and~\ref{eqn:Poisson}. In what follows, we work in terms of $G$ and $\nu$ and further set $G=1$ as our stress unit. In $d=2$ dimensions, we then have $\mu=1$, $K=(1+\nu)/(1-\nu)$ and $\lambda=2\nu/(1-\nu)$. 

We assume the total stress $\Sigma_{ij}(r_i,t)$ at each lattice site to comprise the sum of the elastoplastic contribution $\sigma_{ij}$ just defined, plus a dissipative contribution of viscosity $\eta$,
\begin{equation}
\label{eqn:totalStress}
    \Sigma_{ij}(r_i,t)=\sigma_{ij}(r_i,t)+2\eta D_{ij},
\end{equation}
in which the strain rate tensor in terms of velocity $v_i=\dot{u}_i$ is
\begin{equation}
D_{ij}=\dot{\epsilon}_{ij}=\tfrac{1}{2}\left(\partial_iv_j+\partial_jv_i\right),
\end{equation}
Force balance in the inertialess limit requires that
\begin{equation}
\label{eqn:forceBalance}
0_i=\partial_j\Sigma_{ij}.
\end{equation}

So far, we have considered only a linear elastic solid in tandem with a linear dissipative stress. We now incorporate plasticity, which naturally also introduces nonlinearity. To do so, we assume that when any element's elastic energy $E=\tfrac{1}{2}\sigma_{ij}\epsilon_{ij}$ 
exceeds a threshold $E_{\rm y}$, it yields plastically on the stochastic timescale $\tau_0=1$ (setting our time unit). We set $E_{\rm y}=G/2=1/2$ (in our units), such that elements fail at a typical shear strain $O(1)$. The element's elastic coefficients  $\mu$ and $\lambda$ (and so also  $G$ and $K$) are then set to zero for all subsequent times, such that the element has permanently failed, capturing a local breakage event in our fibre network example. Over a timescale set by $\eta$ divided by the elastic constants, stress then propagates to recover force balance.

In setting $E_{\rm y}=G/2=1/2$, we have chosen the local yield energy $E_{\rm y}$ to scale in proportion to the shear modulus $G$ rather than the bulk modulus $K$. This is because we are investigating a material’s toughness to imposed changes in shape (shear strain) rather than imposed volumetric compression (bulk strain). Indeed, this is the regime relevant to most applications in both nature and technology: stents, bones, blood vessels, textiles, protective pads, helmets etc. typically suffer bending, shearing or stretching (shape changes) in practical usage~\cite{ali2014auxetic,gatt2015hierarchical,evans2000auxetic,sanami2014auxetic,sloan2011helical}, rather than imposed overall bulk volumetric changes, consistent with the classification of fracture modes I-III (opening, in-plane shear and out-of-plane shear) in fracture mechanics~\cite{gdoutos2020fracture}.

Were we to have made the alternative choice, setting $E_{\rm y}$ proportional to $K$ and exploring the auxetic regime of small $K/G$, this would correspond to a material with a very small ratio $E_{\rm y}/G$. Such a material would be very stiff against changes in shape, in having a large shear modulus $G$;  but  also very weak to changes in shape, in having a small yield energy $E_{\rm y}$. Such very stiff, very weak materials would fail at tiny shear strains, even in the absence of any stress propagation and associated spatially cooperative plasticity and fracture propagation.

Indeed, when considering the underlying mesoscopic structure of conventional versus auxetic solids -- solid foam cells with positive versus negative internal angles respectively, for example -- we recognise that the energy scale of local plastic failure events will be set in both cases by the energy involved in locally breaking this mesoscopic structure. For the pedagogically illustrative case of a 2D foam, for example, such events involve breaking the filaments that constitute the foam's cell boundaries. The energy cost of breaking these filaments is expected to be largely independent of the sign of the cell's internal angles (their failure happens at the microscopic lengthscale of the filament cross section, smaller than the mesoscopic scale of the cells that the filaments form), and is therefore the same for both conventional and auxetic materials. Accordingly, we indeed assume that the energy scale $E_{\rm y}$ for local mesoscopic failure events remains $O(1)$ even as the auxetic limit is approached, $K\to 0$.

Details of our numerical algorithm used to simulate this model are given in the methods section below.

\section*{Sample preparation and shear protocol}
\label{sec:preparation}

The importance of initial sample preparation or annealing and disorder to the failure properties of amorphous materials when they are subsequently deformed is increasingly being appreciated. Typically, materials with initially narrower (less disordered) distributions of local strains show more brittle and less ductile failure behaviour, at fixed sample size. 
%In our work, the temperature $T_0$ a material is %initially prepared at prior to any shear being %applied controls the degree of annealing, with %lower $T_0$ giving stronger annealing. 
%To model a sample's equilibration at temperature %$T_0$, 

To model a material's initial state, we note that any element's elastic energy can be written as the sum of shear and compressional contributions:
\begin{equation}
E=G\left(\epsilon_{ij}\epsilon_{ij}-\tfrac{1}{2}\epsilon_{ll}^2\right)+\tfrac{1}{2}K\epsilon_{ll}^2=\tfrac{1}{2}G\left(a-b\right)^2+\tfrac{1}{2}K\left(a+b\right),
\end{equation}
in which we have set $\epsilon_{11}=a, \epsilon_{22}=b$ and $ \epsilon_{12}=\epsilon_{21}=0$ in the diagonal frame of the element's local strain tensor. For simplicity we assume that $\mu$ and $\lambda$ (and so also $G$ and $K$) are initially uniform across a freshly prepared sample, the same for all elements, prior to any later elemental failure events when the sample is sheared. We further assume energy 
equipartition between shear and extension, such that:
\begin{equation}
    \frac{1}{2}G\overline{(a-b)^2}=\frac{1}{2}K\overline{(a+b)^2}\equiv\frac{1}{2}Gl_0^2,
    %k_{\rm B}T_0,
\end{equation}
in which overline denotes averaging over all elements on the lattice. To initialise the elemental strains, therefore, we choose each element's value of $a-b$ (resp. $a+b$) from a Gaussian of variance $l_0^2$ (resp. $Gl_0^2/K$),
%$k_{\rm B}T_0/G$ (resp. $k_{\rm B}T_0/K$). 
%Setting $k_{\rm B}T_0/G\equiv l_0^2$, 
which defines the standard deviation of initial local strains $l_0$. Lower $l_0$ corresponds to a less disordered (or better annealed) sample initially. We then rotate each element's strain tensor through a random angle to give its strain  tensor in the $xy$ frame of the lattice, before
%finally quench to temperature $T=0$ and 
evolving the system to a steady  force balanced state. 

A global shear of slow rate $\gdot\ll 1$ is then imposed for all subsequent times $t>0$ (defining our origin of time), giving an affine contribution  $D_{xy}=D_{yx}=\gdot$ to the strain rate tensor, with Lees-Edwards periodic boundary conditions.  Any element then loads elastically as a function of increasing time $t$, partly via this affine contribution, and partly via the elastic propagation of stress from plastic events that occur elsewhere in the sample, as considered in the next section, until the element itself finally yields plastically. 

To map out phase behaviour in full generality -- as a function of both the degree of initial sample disorder, as characterised by $l_0$, and the degree of which the material is auxetic, as characterised by $\nu$ -- we  assume  that $l_0$ and $\nu$ can be prescribed independently of each other. 

\section*{Reduced stress propagation in auxetic materials}

Following Eshelby~\cite{eshelby1957determination}, Picard et al.~\cite{picard2004elastic} considered the global elastic propagation of stress following a single local plastic event in an incompressible linear elastic material with a homogeneous shear modulus $\mu$. The relevant Green's function is the Oseen tensor $O_{ij}^{\rm I}=\frac{1}{\mu k^2}(\delta_{ij}-\hat{k}_i\hat{k}_j)$, in Fourier space. (We use superscript I to denote incompressible.) Here we generalise Picard's calculation to the full range of material types from incompressible, $\nu=+1$, to maximally auxetic, $\nu=-1^+$, by  considering the elastic propagation of stress following a local plastic event in a linearly elastic material with generalised, but still spatially homogeneous, Lam\'{e} coefficients $\mu$ and $\lambda$.  

\begin{figure}[!t]
  \includegraphics[width=\columnwidth]{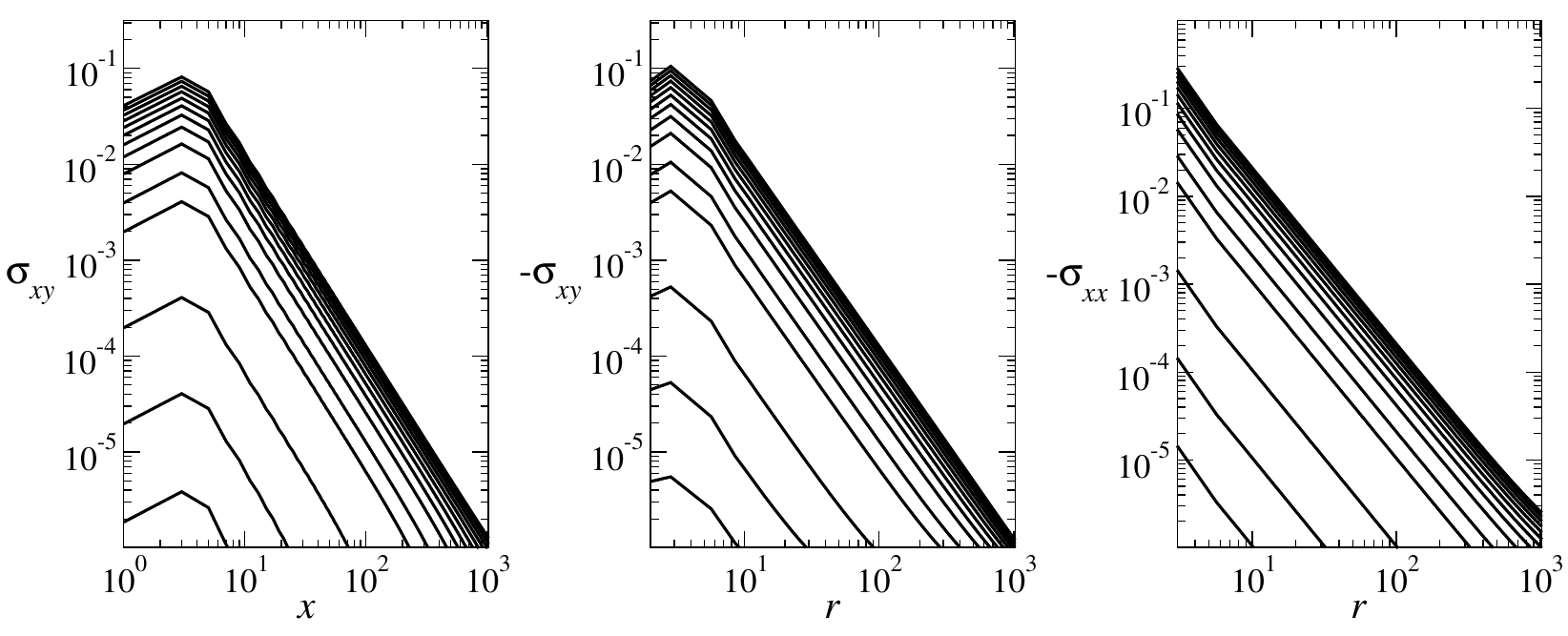} 
  \caption{Elastic stress response to a unit source plastic shear stress relaxation $s_{xy}=S=-1$ at the origin $x=y=0$.  {\bf Left:} elastic shear stress along the centreline $y=0$.  {\bf Middle:} negative of the elastic shear stress along the diagonal $x=y$, with $r=\sqrt{2}x$. {\bf Right:} negative of the elastic normal stress $-\sigma_{xx}=-\sigma_{yy}$ along the diagonal. ($\sigma_{xx}=\sigma_{yy}=0$ along the  centreline.)  Poisson ratio  %$\nu=0.9999,0.8,0.6,0.4,0.2,0.0,-0.2,-0.4,-0.6,-0.8,-0.9,-0.99,-0.999,-0.9999$ 
$\nu=0.9999,0.8,0.6,0.4\cdots -0.6,-0.8,-0.9,-0.99,-0.999,-0.9999$ 
  in curves with peak heights downwards in each panel.  Lattice size $N^2=2048^2$.} 
  \label{fig:propagator} %propagators/propagator_N512_nu{0.9999,0.8,0.6,0.4,0.2,0.0,-0.2,-0.4,-0.6,-0.8,-0.9,-0.99,-0.999,-0.9999}.dat
\end{figure}

Note, however, that in the full elastoplastic model as described in the previous section, and numerically simulated in subsequent sections, after the plastic yielding of an element at any lattice site, $\mu$ and $\lambda$ are set to zero at that site for all subsequent times, to model the permanent breaking of that element. Accordingly, any subsequent plastic events occur in a medium where $\mu$ and $\lambda$ are actually heterogeneous. The calculation outlined in this section thus only strictly applies to the  propagation of stress following the first local plastic event in a freshly prepared material.  Nonetheless, it should provide a reasonable guide to the early stages of material yielding, when few local failure events have yet occurred.

Once the material has reached a new state of force balance, with the propagated elastic stress  balancing the localised plastic stress source, which we denote $s_{ij}$, the total stress $\sigma_{ij}=2\mu\epsilon_{ij}+\lambda\epsilon_{ll}\delta_{ij}-s_{ij}$
%
%\begin{equation}
%\label{eqn:stressTotal}
%\sigma_{ij}=2\mu\epsilon_{ij}+\lambda\epsilon_{ll}\delta_{ij}-s_{ij},
%\end{equation}
%
obeys the condition of force balance
\begin{equation}
\label{eqn:forceBalance1}
0_i=\partial_j\sigma_{ij}=2\mu\partial_j\epsilon_{ij}+\lambda\partial_j\epsilon_{ll}\delta_{ij}-\partial_j s_{ij}.
\end{equation}
Substituting into this the expression for strain from Eqn.~\ref{eqn:strain},
and transforming from real to Fourier space $r_i\to k_i$, we obtain 
\begin{equation}
-ik_js_{ij}=\mu(k_ik_ju_j+k_j^2u_i)+\lambda k_ik_ju_j.
\end{equation}
Rearranging gives the elastic displacement response 
\begin{equation}
u_i=-O_{ij}f_j,
\end{equation}
to the stress source $s_{ij}$ and corresponding force $f_i=ik_js_{ij}$. The propagation tensor
\begin{equation}
O_{ij}=-\tfrac{1}{2}(1+\nu)O_{ij}^{\rm I}-\tfrac{1}{2}(1-\nu)O_{ij}^{\rm A},
\end{equation}
in which
\begin{equation}
O_{ij}^{\rm I}=\frac{1}{\mu k^2}(\delta_{ij}-\hat{k}_i\hat{k}_j)\;\;{\rm and}\;\;O_{ij}^{\rm A}=\frac{1}{\mu k^2}\delta_{ij}.
\label{eqn:Oseen}
\end{equation}

For an incompressible material $\nu\to 1$, this recovers Picard's result 
$O_{ij}=O_{ij}^{\rm I}=\frac{1}{\mu k^2}(\delta_{ij}-\hat{k}_i\hat{k}_j)$. For a maximally auxetic material, $\nu\to -1^+$, we instead have $O_{ij}=O_{ij}^{\rm A}=\frac{1}{\mu k^2}\delta_{ij}$, with superscript A denoting auxetic. At intermediate $\nu$ we have a linear interpolation between these two limiting cases.

The associated strain and stress fields are then obtained by substituting this result for $u_i$ into Eqns.~\ref{eqn:strain} and~\ref{eqn:stress1}. 
For any deformation protocol in which shear stresses dominate, of particular interest is  the elastic shear stress response $\sigma_{xy}$ to a plastic shear stress source 
$s_{ij}=S(\delta_{ix}\delta_{jy}+\delta_{iy}\delta_{jx})$ localised at the origin in real space. In Fourier space this is given by
\begin{equation}
\label{eqn:propagator}
    \sigma_{xy}=S\left[1-2(1+\nu)\frac{k_x^2k_y^2}{k^4}\right].
\end{equation}

This quantity is plotted in real space along the centreline $y=0$ and diagonal $x=y$ in the left and middle panels of Fig.~\ref{fig:propagator}. 
In the incompressible limit, $\nu=1$, we recover Picard's result, with a stress propagator that is quadrupolar in angle and decays with distance as  $\sim 1/r^2$. In important contrast, as $\nu$ decreases towards the maximally auxetic limit $\nu=-1^+$,  the quadrupolar $1/r^2$ contribution  decreases linearly to zero: in a fully auxetic material, a local plastic relaxation of shear stress has only a {\em local} consequence at the site where relaxation occurred, with no propagation across the material. Indeed, the maximum value in each panel of Fig.~\ref{fig:propagator} decreases linearly to zero as $\nu\to -1$. The normal stress  component $\sigma_{xx}=\sigma_{yy}$ is shown along the diagonal $x=y$ in the right panel of Fig.~\ref{fig:propagator}. (It is zero along the centreline.)

\section*{Increased failure toughness of auxetic materials}

\begin{figure}[!t]
 \includegraphics[width=1.0\columnwidth]{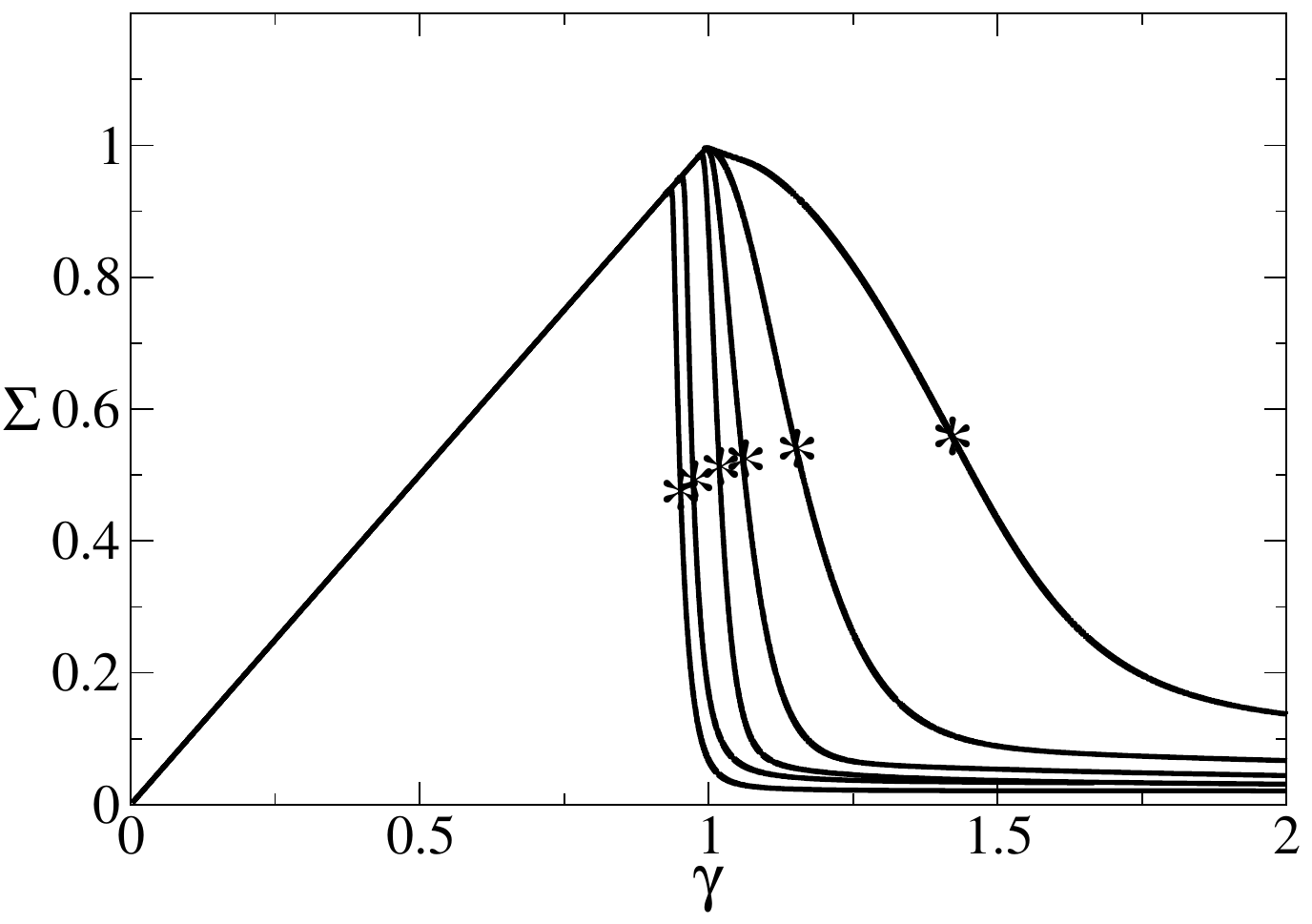}
  \caption{Shear stress versus shear strain for Poisson ratio values $\nu=0.99,0.00,-0.90,-0.99,-0.999,-0.9999$
  in curves left to right. Strain rate $\gdot=10^{-3}$, lattice size $N^2=512^2$, standard deviation of initial strain values $l_0=0.025$, timestep $\alpha=0.003$. Each curve represents a single run, with no averaging over random number seed. Yield strains $\gamma^*$ indicated by asterisks.} 
  \label{fig:stressStrain}
  %auxeticAFSbreak3.c_fastSlow1_balance1_gdot1.0e-3_nu{0.99,0.00,-0.90,-0.99,-0.999,-0.9999}_l00.025_eta1.0_Lx1.0_Ly1.0_Nx512_Ny512_Dt0.003_gammax2.0_seed3
\end{figure}

In the previous section, we considered analytically the elastic propagation of stress following a single local plastic relaxation event in a material with homogeneous Lam\'e coefficients. We now numerically simulate our full elastoplastic model in which all $N\times N$ elastoplastic elements interact. Each element now experiences affine loading at the imposed shear rate $\gdot$, additional elastic deformation as a result of stress propagation from the plastic failure of elements elsewhere in the sample, and its own eventual local plastic failure. As noted above, once any element fails, its Lam\'{e} coefficients are set to zero and the material's elastic constants become heterogeneous.

Fig.~\ref{fig:stressStrain} shows the shear stress $\Sigma_{xy}\equiv \Sigma$ as a function of the globally imposed shear strain $\gamma=\gdot t$, for a well prepared sample with a small standard deviation of initial strains. Curves left to right correspond to decreasing values of the Poisson ratio $\nu$. The leftmost curve is for a nearly incompressible material and the rightmost curve is for an almost maximally auxetic material. As can be seen, in moving from the incompressible to auxetic case, the strain at which failure occurs increases: i.e., the material's mechanical failure resistance increases.

\begin{figure}[!t]
 \includegraphics[width=1.0\columnwidth]{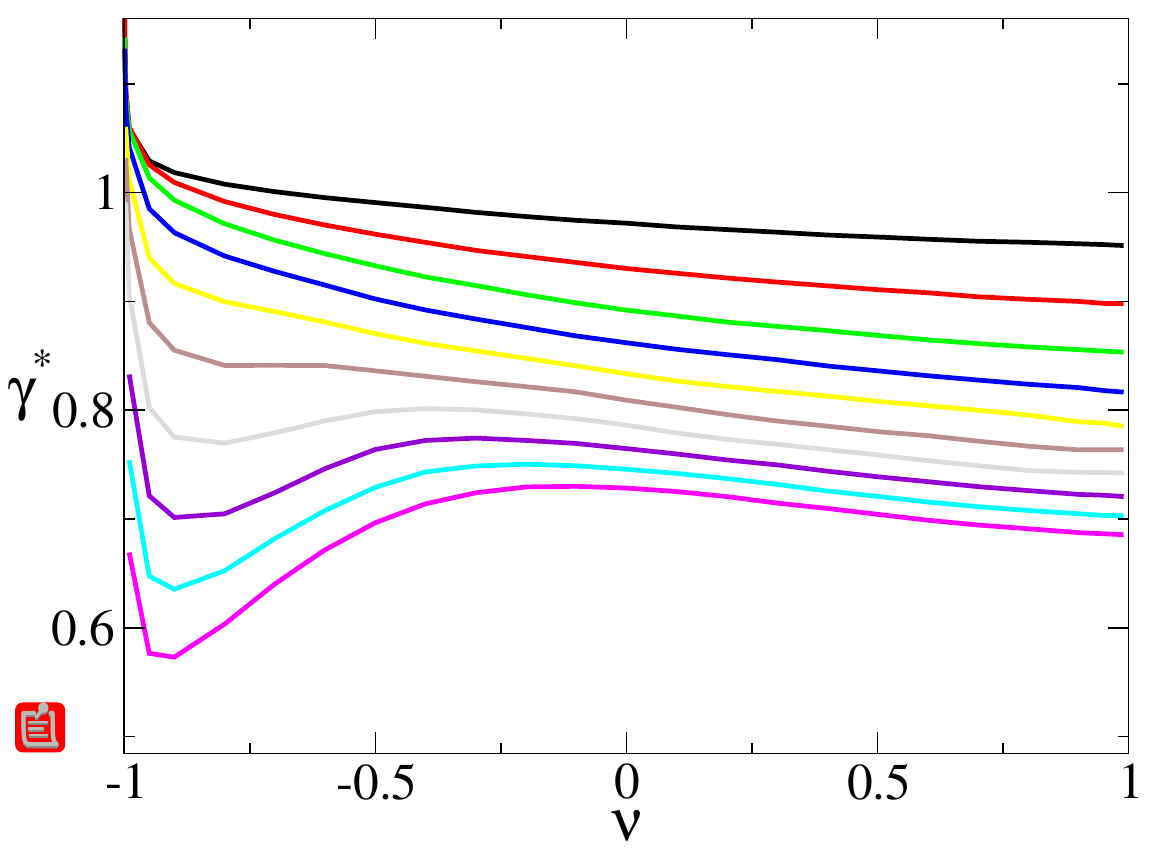}
  \caption{Yield strain as a function of Poisson ratio $\nu$. Curves downwards correspond to increasing values of the standard deviation of initial strain values $l_0=0.025,0.050\cdots 0.250$. Each data point is averaged over runs performed at ten different values of the random number seed. Strain rate $\gdot=10^{-3}$, lattice size $N^2=512^2$, timestep $\alpha=0.003$.} 
  \label{fig:yieldStrain}
  %auxeticAFSbreak3.c_fastSlow1_balance1_gdot1.0e-3_nu*_l0*_eta1.0_Lx1.0_Ly1.0_Nx1024_Ny1024_Dt0.003_gammax2.0_seed{1-10}
\end{figure}

To quantify this, we define the yield strain $\gamma^*$ for any sample as the strain at which the stress falls half way from its peak value to its eventual value at the maximum strain simulated, $\gamma=2.0$. For each curve in Fig.~\ref{fig:stressStrain}, this yield strain $\gamma^*$ is indicated by an asterisk. This quantity $\gamma^*$ is then plotted as a function of Poisson ratio $\nu$ for several different levels of disorder in the initial strain distribution in Fig.~\ref{fig:yieldStrain}. 
For well annealed samples, with a small standard deviation of initial local strain values $l_0$ (top curve), the yield strain increases monotonically in moving from the incompressible case $\nu=1.0$ towards the maximally auxetic limit, $\nu=-1.0^+$. In contrast, for initially disordered samples with a large standard deviation of initial local strains (bottom curve), the yield strain first increases with decreasing $\nu$, then decreases, before increasing again as the auxetic limit is approached, $\nu\to -1^+$. We return to comment further on this non-monotonic dependence below.

\begin{figure*}[!t]
 \includegraphics[width=\textwidth]{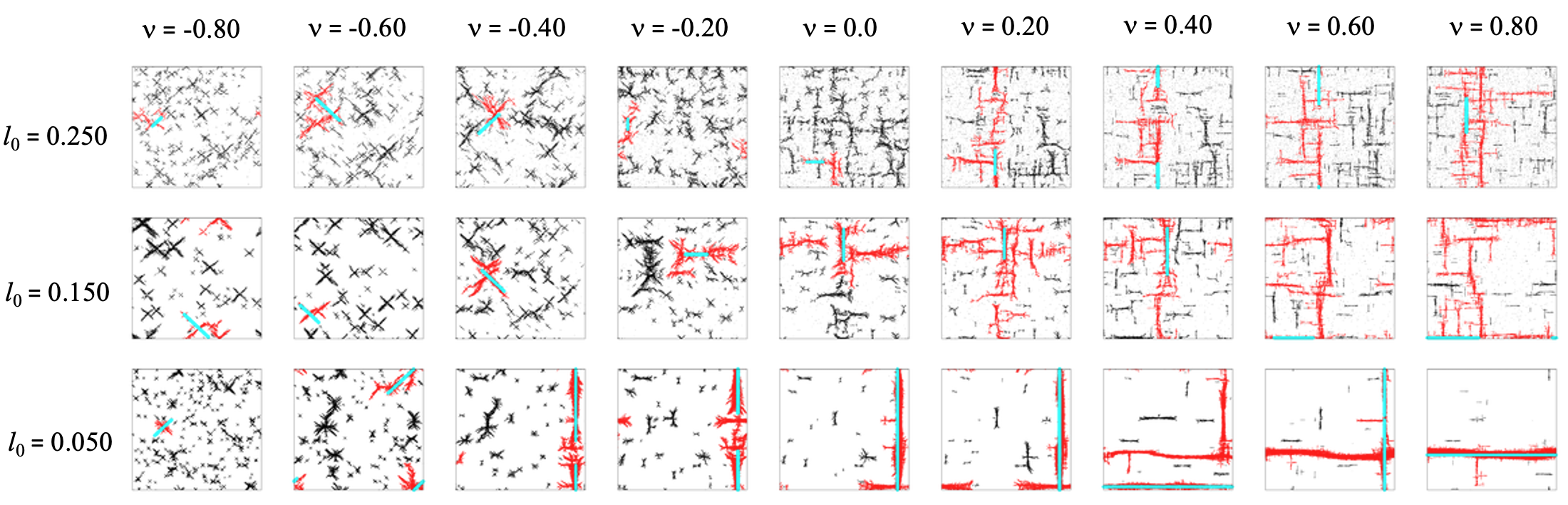}
  \caption{State snapshots showing the sites that have plastically failed by the time the yield strain $\gamma^*$ is attained. Poisson ratio $\nu=-0.80,-0.60,-0.40,-0.20,0.00,0.20,0.40,0.60,0.80$ in columns left to right. Standard deviation of initial local strain values $l_0=0.050, 0.150, 0.250$ in rows from bottom to top. Any site that has failed is shown in either black, red or cyan. Red dots indicate the sites that belong to the largest areal cluster of adjacent sites to have failed. Cyan lines indicate the sites within each largest areal cluster that belong to the longest linear array of adjacent sites to have failed, either along $x$ or $y$ or the diagonal $x=y$ or $x=-y$. Black dots indicate sites that have failed but do not belong to the largest cluster.} 
  \label{fig:states}
  %auxeticAFSbreak3.c_fastSlow1_balance1_gdot1.0e-3_nu*_l*_eta1.0_Lx1.0_Ly1.0_Nx1024_Ny1024_Dt0.003_gammax2.0_seed1
\end{figure*}

So far, we have discussed the macroscopically measured mechanical shear stress signal, and the corresponding  macroscopic yield strain $\gamma^*$. Perhaps more important than these macroscopic measures, however, is the spatial patterning of plastic damage that accumulates internally as a function of position across the material. To investigate this, we plot in Fig.~\ref{fig:states} snapshots showing the sites that have undergone local plastic yielding by the time the yield strain $\gamma^*$ is reached. Snapshots are shown on a grid of values of Poisson ratio $\nu$ (left to right) and initial strain disorder $l_0$ (bottom to top). 


\begin{figure}[!b]
 \includegraphics[width=1.0\columnwidth]{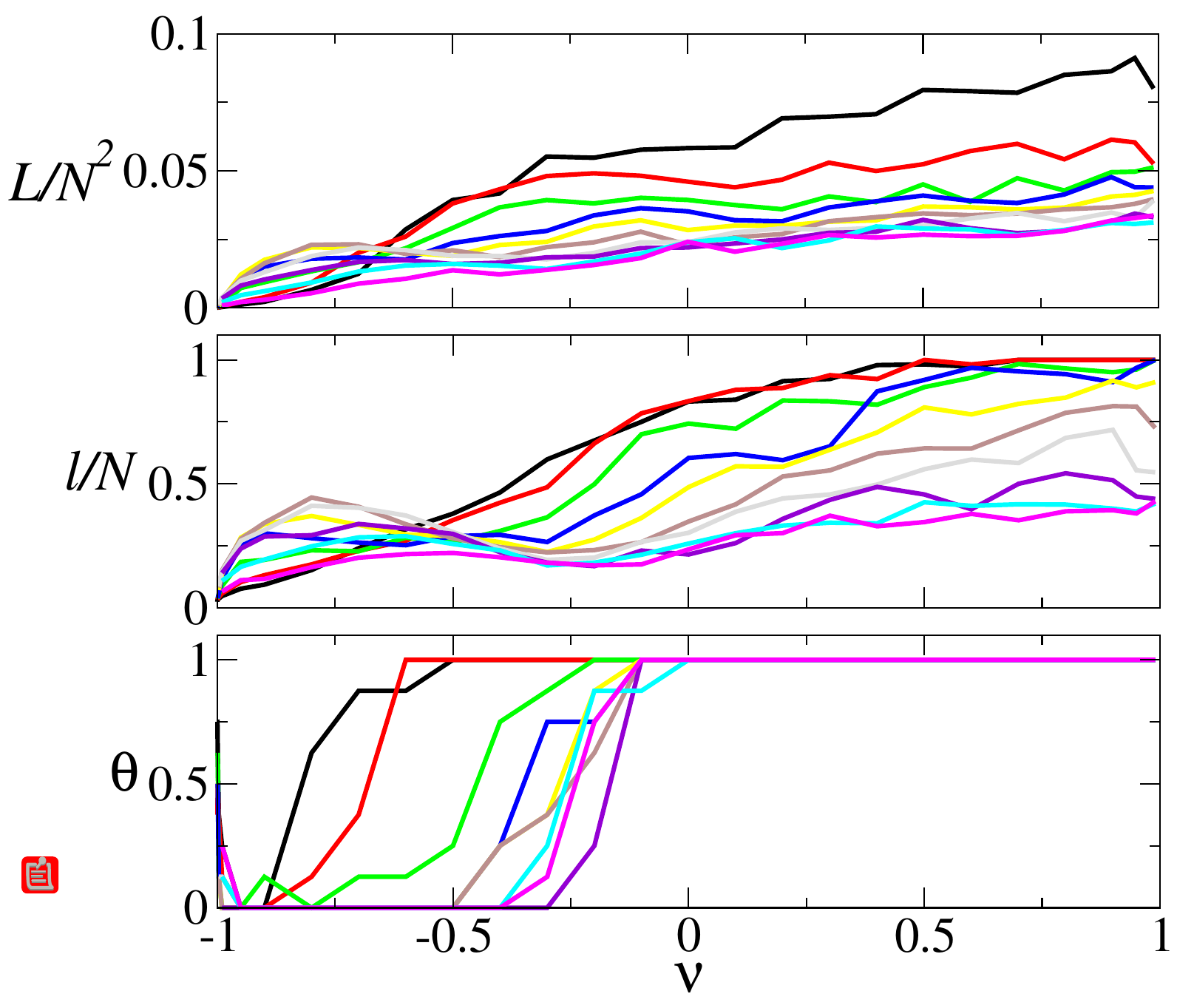}
  \caption{{\bf Top:} areal size of the largest cluster of adjacently failed sites at yield. {\bf Middle:} linear length of the longest line of adjacently failed sites at yield, either along $x$ or $y$ or $x=y$ or $x=-y$. {\bf Bottom:} orientation of this longest line, with $1$ corresponding to cracks along $x$ or $y$, and $0$ corresponding cracks along $x=y$ or $x=-y$. Each quantity is plotted as a function of the Poisson ratio $\nu$. The standard deviation of initial strain values $l_0=0.025,0.050\cdots 0.250$ in black, red, green, blue, yellow, brown, grey, violent, cyan and magenta curves respectively. Each data point is averaged over runs performed at ten different values of the random number seed.
   Strain rate $\gdot=10^{-3}$, lattice size $N^2=512^2$,  timestep $\alpha=0.003$.} 
  \label{fig:stats}
  %auxeticAFSbreak3.c_fastSlow1_balance1_gdot1.0e-3_nu*_l0*_eta1.0_Lx1.0_Ly1.0_Nx1024_Ny1024_Dt0.003_gammax2.0_seed{1-10}
\end{figure}

For each pairing of values of $\nu$ and $l_0$, we identify in red the {\em largest areal cluster} of adjacently failed sites, and shall denote the number of sites in this cluster by $A$. (Before identifying clusters, we in fact coarse grain each site with its immediately adjacent neighbours. We do so because successive plastic yielding events can often jump by two sites due to the propagator being roughly equal between the first and second sites adjacent to a local failure.) For each such largest cluster, we then identify in cyan the {\em longest linear array} of adjacently failed sites (after coarse graining by one, as just described), either along  $x$ or $y$ or the diagonal $x=y$ or $x=-y$. We denote the number of failed sites along this line by $l$. For each such longest line, we further define an angular variable $\theta$, which adopts the value $1$ if the line is along either $x$ or $y$ (each case being equally likely, due to the quadrupolar nature of the stress propagator), and a value $0$ if the line is along $x=y$ or $x=-y$ (each again equally likely for the same reason). Each of these three measures  -- the areal size $A$ of the largest plastic cluster at yield, its longest linear length $l$, and orientation $\theta$ -- is plotted as a function of Poisson ratio in Fig.~\ref{fig:stats}, for several different levels of disorder in the initial strain distribution. 

Taking Figs.~\ref{fig:stats} and~\ref{fig:states} together, several observations are notable.  First, for incompressible materials $\nu\to 1$ (right of Fig.~\ref{fig:states}), and for materials with only a low level of disorder in the initial local strain distribution (bottom row), clusters of local plasticity percolate  across the sample in uninterrupted lines along $x$ or $y$. For incompressible samples that have instead a greater degree of local strain disorder initially (top right snapshots), such clusters still percolate along $x$ or $y$, but do so in a slightly more diffuse and disordered way. Such behaviour has indeed been observed for different levels of initial sample annealing previously in incompressible systems, albeit in the case where elastoplastic elements reform after failure~\cite{pollard2022yielding,lin2015criticality}. Either case corresponds to macroscopic material failure, with a line of plastic damage that percolates across the sample. This percolation arises from the spatial stress propagation considered in the previous section: when an element yields at any site, sites that are nearby along $x$ and $y$ suffer increased loading, and fail in turn.

As we move leftwards across Fig.~\ref{fig:states} and the Poisson ratio decreases from its value $\nu=1$ for an incompressible material, towards the limit of a perfectly auxetic material, $\nu\to -1^+$, the size of the largest cluster of plastically failed sites decreases towards zero (top and middle panels of Fig.~\ref{fig:stats}). This is true regardless of the level of disorder in the initial strain distribution and is consistent with  our analytical calculation of the stress propagator above: in the auxetic limit $\nu\to -1^+$, there is no spatial propagation of stress and a  local plastic failure at any site has no consequence any for other site within the material. Spatially correlated cracks of plasticity therefore cannot form. For $\nu$ values close to the auxetic limit $-1^+$, although a material may still be judged to have yielded according to the fall in the externally measured stress signal, any internal plastic damage is dispersed diffusely across the material in small microcracks, without catastrophic macroscopic failure. 

Just as the size of the largest cluster of local plasticity decreases with decreasing Poisson ratio, its angular orientation also changes. For an incompressible material, $\nu\to 1$, the largest cluster of adjacent plastic events is always either along $x$ or $y$. In contrast, for materials close to the auxetic limit, $\nu\to -1^+$, the clusters  instead propagate diagonally along $x=y$ or $x=-y$. This is seen in Fig.~\ref{fig:states}, moving leftward along each row, and via the dependence of $\theta$ on $\nu$ in Fig.~\ref{fig:stats} (bottom).

\section*{Discussion}

We have studied theoretically the increased mechanical failure resistance of auxetic materials compared with conventional ones. Indeed, in a conventional incompressible amorphous material, an initially localised microscopic plastic event causes a propagation of stress that then triggers knock-on failure events in neighbouring regions of the material. This in turn leads to a runaway instability, with a spatio-temporal cooperation of many such events on large lengthscales (recall Fig.~\ref{fig:states} for materials with high Poisson ratio $\nu$) and  a rapid drop in stress (recall Fig.~\ref{fig:stressStrain} for high $\nu$). This sudden brittle material failure and fracture has been the focus of intense study in conventional amorphous materials ~\cite{rozen2020fast,barlow2020ductile,pollard2022yielding,rossi2022finite,ozawa2018random,popovic2018elastoplastic}. 

In this work, in contrast, we have shown analytically that the propagation of stress following a localised plastic event is reduced with decreasing Poisson ratio $\nu$, with no non-local stress propagation in the auxetic limit $\nu\to-1^+$. By numerically simulating a simple lattice elastoplastic model that couples many elastoplastic elements together, we further showed that this reduced spatial stress propagation leads to an increased mechanical failure resistance in auxetic materials, as characterised both by an increased macroscopic strain at which yielding occurs, and a decreased clustering of plastic failure events across a material.

As noted above, this increased resistance of auxetics to mechanical failure has been demonstrated in experimental studies in several different classes of auxetic materials~\cite{hall2008sign,yang2017fracture,donoghue2009fracture,choi1996fracture,bezazi2009tensile,lakes1987foam,lakes1993indentability}. Typical negative Poisson ratios achieved experimentally range from around $-0.15$ for laminates~\cite{donoghue2009fracture}, $-0.20$ for carbon nanotube sheets~\cite{hall2008sign}, $-0.54$ for auxetics obtained by elastic buckling~\cite{bertoldi2010negative}, $-0.60$ for those obtained by bond pruning of random networks~\cite{reid2018auxetic} to $-0.7$ for flexible polymer foams and $-0.8$ for metal foams~\cite{choi1996fracture,lakes1993indentability}. 
In apparent contradiction, however, the fracture energy of metallic glasses was found to {\em decrease} with decreasing Poisson ratio in Refs.~\cite{lewandowski2005intrinsic,greaves2011poisson}. It should be noted, however, that only positive (non-auxetic) values of Poisson ratio pertained in those materials. Future work is nonetheless warranted to resolve this apparent contradiction. 

This work has also revealed  a change in orientation of clusters of plasticity as the Poisson ratio decreases, from cracks along $x$ or $y$ in incompressible materials to diagonal cracks along $x=y$ or $x=-y$ in auxetic ones. (Recall Fig.~\ref{fig:states}.) This can be understood  by recalling that cracks form via a plastically yielding site giving a stress perturbation $d\sigma_{xy}$ to its neighbours, and so on in a repeating way across the lattice.  When any element at some origin site indeed yields, it relaxes all its stress components, both shear and normal. The spatially propagated $d\sigma_{xy}$ that results from the relaxation of shear stress at this origin site has already been considered in Eqn.~\ref{eqn:propagator}: it is quadrupolar, dominating along the horizontal and vertical axes to  give the patterns of cracking seen for larger $\nu$ values in Fig.~\ref{fig:states}. This term however scales as $1+\nu$ and becomes small in the auxetic limit $\nu\to -1$. 
For the smallest (most negative) $\nu$ in Fig.~\ref{fig:states}, the spatially propagated $d\sigma_{xy}$ instead predominantly stems from the relaxation of normal stress at the origin site. Considering a normal stress source $s_{ij}$ in Eqns.~\ref{eqn:forceBalance1} to ~\ref{eqn:Oseen}, it is easily shown that the propagated stress $d\sigma_{xy}$ is largest along the diagonals, consistent with the diagonal cracks observed for the smaller values of $\nu$ in Fig.~\ref{fig:states}.

% Note - site that yields: epsxy = O(1) by advection. epsxx, epsyy suffer no advection, so when they yield their standard deviation is as initialised, it l0/sqrt(delta) where delta=1+nu. So sigxy=O(1), sigxx, sigyy = l0*sqrt(delta). relaxing sigxy propagates to dsixy in vicinity via O^I which is O(1), giving quadrupolar change dsigxy strongest along horizontal and vertical. relaxing sigxx and sigyy propagate to dsigxy in vicinity via O^A giving dipolar term O(sqrt(delta l0)) strongest along the diagonals. This explains the crossover dependence on delta and l0 seen in fig 4

%Our work has also revealed a change in orientation %of clusters of plasticity as the Poisson ratio %decreases, from cracks along $x$ or $y$ in %%incompressible materials to diagonal cracks along %$x=y$ or $x=-y$ in auxetic ones. 
Preliminary observations suggest that the same competition between shear and normal stresses underpins the non-monotonic dependence of yield strain on Poisson ratio for samples with a high degree of disorder in the initial distribution of local strains. Future work should study these phenomena more fully.  Indeed, in the crossover regime between the two different dominant cluster orientations (along $x$ or $y$; or along $x=y$ or $x=-y$), increasingly intricate clusters of plasticity are seen in Fig.~\ref{fig:states}, with competing branches in the different orientations. A careful study with increasing system size would be warranted, to establish whether infinitely branched fractal clusters may arise in some regimes of Poisson ratio.  Given the potentially increasing importance of normal stresses at low values of the Poisson ratio, it would also be interesting to consider different possible constitutive models of normal vs shear stresses. 

\section*{Numerical methods}

On each site of a $d=2$ dimensional $N\times N$ lattice, we initialise a single elastoplastic element as described in the section `sample preparation' in the main text. We then use a time-stepping algorithm to shear the material, as follows. At each timestep, we execute three sub-steps. The first comprises an elastic update, in which the elastoplastic shear strain of each element is incremented by the affine shear rate multiplied by the timestep, and the change in shear strain used to compute the corresponding change in shear stress via linear elasticity with the appropriate Lam\'{e} coefficients. The second substep comprises a plastic update, in which each element is tested for whether it exceeds the threshold for local plastic yielding. Any element that does is then yielded with a probability given by the timestep divided by the time constant $\tau_0$. When any element yields, its Lam\'{e} coefficients are set to zero. The elastoplastic stress field is then recalculated. In the third sub-step, the elastoplastic stress field is transformed to Fourier space, using periodic boundary conditions, and the non-affine velocity field is calculated via equations~\ref{eqn:totalStress} to~\ref{eqn:forceBalance}. From this non-affine velocity field the non-affine strain rate field is calculated and multiplied by the time-step to give the non-affine strain update, which is added to the elastoplastic strain. At each step, the numerical timestep is set adaptively via a scale factor $\alpha$ divided by the maximum dynamical rate in the system.

The author thanks Mike Cates and Peter Sollich for discussions. This project has received funding from the European Research Council (ERC) under the European Union's Horizon 2020 research and innovation programme (grant agreement No. 885146). 

%\bibliography{auxetic.bib}

%apsrev4-2.bst 2019-01-14 (MD) hand-edited version of apsrev4-1.bst
%Control: key (0)
%Control: author (8) initials jnrlst
%Control: editor formatted (1) identically to author
%Control: production of article title (0) allowed
%Control: page (0) single
%Control: year (1) truncated
%Control: production of eprint (0) enabled
%

\end{document}